\documentclass[fleqn,usenatbib]{mnras}
\usepackage{newtxtext,newtxmath}
\usepackage{amssymb}
\usepackage{graphicx}
\usepackage{epsfig}

\newcommand{\chandra}{{\it Chandra}}

\newcommand{\xmm}{{\it XMM-Newton}}

\newcommand{\fermi}{{\it Fermi}}
\newcommand{\nustar}{{\it NuSTAR}}
\newcommand{\msp}{PSR\,J2129--0429}

\title[High-Energy Observations of \msp]{Broadband High-Energy Emissions of the Redback Millisecond Pulsar PSR J2129--0429}

\author[A.~K.~H.~Kong et al.]{Albert~K.~H.~Kong$^1$\thanks{E-mail: akong@gapp.nthu.edu.tw}, Jumpei Takata$^2$, C.~Y. Hui$^3$, J. Zhao$^2$, K.~L. Li$^4$ and \newauthor P.~H.~T. Tam$^5$\\
$^1$Institute of Astronomy, National Tsing Hua University, Hsinchu 30013, Taiwan\\
$^2$School of Physics, Huazhong University of Science and Technology, Wuhan 430074, China\\
$^3$Department of Astronomy and Space Science, Chungnam National University, Daejeon, Republic of Korea\\
$^4$Department of Physics and Astronomy, Michigan State University, East Lansing, MI 48824, USA\\
$^5$School of Physics and Astronomy, Sun Yat-sen University, Zhuhai 519082, China
}

\begin{document}
\label{firstpage}
\pagerange{\pageref{firstpage}--\pageref{lastpage}}
\maketitle

\label{firstpage}

\begin{abstract}
We present the first results from a joint \xmm\ and \nustar\ observation of the gamma-ray emitting millisecond pulsar compact binary \msp. X-ray emission up to about 40 keV is detected and the joint spectrum can be modeled with a power-law plus a neutron star atmosphere model. At a distance of 1.4 kpc, the 0.3-79 keV luminosity is $3.56\times10^{32}$ erg s$^{-1}$. We also detected the 0.64-day binary orbital period with a double-peaked structure across the wavebands. By combining the updated \fermi\ GeV data, we modeled the broadband spectral energy distribution as well as the X-ray modulation with an intrabinary model involving shock interaction between pulsar wind and outflow from the companion star. Lastly, we reported a high-resolution X-ray image provided by \chandra\ to 
rule out the proposed pulsar wind nebula associated with \msp.
\end{abstract}


\section{Introduction} \label{sec:intro}
A dynamic group of millisecond pulsars (MSPs) which are commonly referred as ``redbacks" has been emerged in the last decade. 
These compact binaries have an orbital period $P_{b}\lesssim20$~hrs. Also, their companions are likely to be 
non-degenerate and have a mass in the range of $M_{c}\sim0.2-0.7M_{\odot}$. A number of interesting phenomena have been observed 
from these objects, which includes: swinging between accretion-powered and rotation-powered states, 
evaporation of an accretion disc, formation of a new disc, intrabinary shocks (see Hui 2014 for a recent review).
These systems provide us with a key to better understand the evolution of low-mass X-ray binaries (LMXBs) and MSPs. 
Furthermore, a recent statistical analysis shows that the X-ray emission of redbacks are generally brighter and 
harder than the other classes of MSPs (Lee et al. 2018, submitted). 
 
Owing to the coordinated efforts of multiwavelength searches, the population of redbacks is growing. So far, 
there are 22 confirmed redbacks have been discovered including 12 in the Galactic field and 10 in globular clusters.
\footnote{For updated statistics, please refer to https://apatruno.wordpress.com/about/millisecond-pulsar-catalogue/} 
A number of redback candidates have also been identified from the unassociated $\gamma-$ray sources 
(e.g., Kong et al. 2014; Hui et al. 2015a; Li et al. 2016). 

Radio pulsation search in the direction towards the $\gamma-$ray source 3FGL J2129.6--0427 has led to the discovery of the 
redback MSP \msp\ (Hessel et al. 2011). This pulsar spins at a periodicity of $P_{s}\sim7.62$~ms resides in a 
binary with an orbital period of $P_{b}\sim0.64$~day (Hessel et al. 2011; Ray et al. 2012; Bellm et al. 2016). 
Its companion has a mass $0.44$~$M_{\odot}$ and has filled up 
its Roche lobe by $\sim95\%$ (Bellm et al. 2016). Since its lower limit on $M_{c}$ is apparently higher than most of the other 
redback MSPs, together with the relatively strong surface magnetic field $B\sim1.6\times10^{9}$G and young characteristic 
age $\tau\sim4\times10^{8}$~yrs, \msp\ is suggested to be in an early stage of recycling (Noori et al. 2017).  
Using an updated electron density model of our Galaxy (Yao et al. 2017), the dispersion measure of PSR J2129--0429 suggests a 
distance of $d\sim1.4$~kpc. For comparison, a distance of $1.8\pm0.1$ kpc is derived from modelling the optical light curves (Bellm et al. 2016). Furthermore, the parallax measurement of the Gaia DR2 data suggests a distance of $2.13^{+0.49}_{-0.35}$ kpc (Bailer-Jones et al. 2018). It is worth nothing that parallax measurement can deviate from the actual distance for MSPs (Igoshev et al. 2016). We will use $d=1.4$ kpc throughout this paper.

Hui et al. (2015b) have carried out a detailed multiwavelength investigation of PSR J2129--0429. Using the data obtained 
by the EPIC camera onboard \emph{XMM-Newton}, a significant X-ray modulation with a double-peaked structure is detected. Its
X-ray emission is non-thermal dominant and can be modeled by a power-law with $\Gamma\sim1.2$. Intrabinary shock is likely to be 
origin of the observed X-rays. The X-ray conversion efficiency in 0.3-10~keV is found to be $L_{x}/\dot{E}\sim0.16\%$. 
Besides the non-thermal shock emission, a thermal component is also suggested to present in all 
orbital phases which can be originated from the heated polar cap resulted from the bombardment of the backflow current. 

Apart from the X-ray, the Optical Monitor onboard \emph{XMM-Newton} also provides the UV light curve of this binary 
(see Fig.~2 in Hui et al. 2015b). A modulation have two peaks with a phase separation of $\sim0.5$ is clearly detected. It can 
be modeled by an ellipsoidal variation with an almost edge-on viewing angle. The heating effect in the UV light curve is 
negligible, consistent with subsequent modeling of optical light curves (Bellm et al. 2016). This indicates the $\gamma-$ray beams of the pulsar is not directed towards the companion. In examining the $\gamma-$ray 
emission in 0.1-300~GeV, Hui et al. (2015b) found that a power-law with an exponential cut-off is preferred over a simple power-law. 
This is similar to the spectral behaviour of the magnetospheric emission as seen in the other pulsars (Abdo et al. 2013). Together 
with the absence of orbital modulation in $\gamma-$rays, this suggests the observed $\gamma-$rays are likely to be originated 
from the pulsar magnetosphere. 

To better constraining the energetics of the intrabinary shock emission, it is important to investigate the system in hard X-ray 
regime (i.e. $\gtrsim10$~keV). A broadband X-ray analysis can allow us to probe if there is steepening in the synchrotron spectrum. 
Such spectral information can be diagonstic for the shock physics (e.g. particle injection rate, magnetic field strength in the 
shock region, acceleration timescale). In this paper, we reports a detailed broadband X-ray analysis of PSR J2129--0429 in the
$0.3-40$~keV band by using the data obtained by \emph{NuSTAR} and \emph{XMM-Newton}. Furthermore, we include over 7 years of \fermi\ GeV data to perform a broadband spectral modeling of \msp.

Apart from examining the X-ray orbital modulation, Hui et al. (2015b) have also identified a putative extended feature associated 
with \msp\ from the \emph{XMM-Newton} image (see Fig.~1 in Hui et al. 2015b). The feature extends for $\sim34$ arcsec toward west 
from the pulsar. This leads to a speculation that such feature is a pulsar wind nebula (PWN) associated with \msp. 
So far, there are only two MSPs, PSR~B1957+20 (Stapper et al. 2003; Huang et al. 2012) and 
PSR~J2124-3358 (Hui \& Becker 2006; Romani et al. 2017; Hui 2018), have their extended X-ray feature detected. 

While the data obtained by \emph{NuSTAR} and \emph{XMM-Newton} can provide tight constraints on the intrabinary shock, 
the angular resolution of these two observatories does not allow us to determine the extended emission is genuine or it 
is originated from any unresolved source. In view of this, we have also observed \msp\ with the \emph{Chandra} 
X-ray Observatory for obtaining imaging data with superior spatial resolution. 

\section{Observations and Data Reduction}

\subsection{NuSTAR}

\msp\ was observed with \nustar\ (Harrison et al. 2013) on 2015 December 23 for $\sim 44$ ks (ObsID 30101057002) and on 2016 April 9 for $\sim 78$ ks (ObsID 30101057004). We downloaded both datasets from the archival database and used the \nustar\ Data Analysis Software (NuSTARDAS) v1.7.0 together with instrumental calibration data files from CALDB version 20170120 to reprocess the raw data.
The calibrated and cleaned event lists were processed with {\tt nupipeline} following the standard procedure. All the data products including images, light curves, spectra and the corresponding response matrices for each of the two focal plane modules (FPMA/B) were generated by the tool {\tt nuproducts}. 
We followed the instructions given in the \nustar\ Data Analysis Software Guide\footnote{http://heasarc.gsfc.nasa.gov/docs/nustar/analysis/nustar\_swguide.pdf} to clean and filter the event lists with standard parameters. Moreover, a barycentric correction was applied to the arrival times of all events based on the \xmm\ position at R.A. (J2000) = 21:29:45.25, Decl. = -04:29:07.95 (Hui et al. 2015) with the tool {\tt barycorr}. In both observations, \msp\ is clearly seen and it is the only source in the field of view. 

Since there is almost no source photons above 40 keV, we limited our data analysis in the energy range of 3--40 keV. Source products including images, light curves, and spectra were obtained from circular regions with a radius of 50 arcsec centered at the position of \msp. An annulus region was employed for background subtraction. The X-ray spectra of the source extracted from FPMA and FPMB were rebinned to have at least 25 counts per bin.

\subsection{XMM-Newton}
Because \nustar\ is not sensitive below 3 keV, we therefore employed \xmm\ data to constrain the spectral shape of soft X-rays. We used the same data products produced in Hui et al. (2015b). In brief, the observation was taken on 2013 October 28 for a total exposure time of $\sim80$ ks. The data reduction procedure can be referred to Hui et al. (2015b).

\subsection{Chandra}
We have observed \msp\ with \emph{Chandra} ACIS-S3 back-illuminated CCD on 2017 September 20 for $\sim50$~ks 
(ObsID: 18991; PI: Hui). 
We used the Chandra Interactive Analysis of Observations software (CIAO) v4.9 with the updated calibration
data base (CALDB version 4.7.7) throughout the analysis. First, we have reprocessed all the data with
subpixel event repositioning in order to facilitate a high angular resolution analysis by using the CIAO script 
{\tt chandra$\_$repro}. To minimizing the background contamination, we restricted our analysis in the energy range 
of $0.3-7$~keV. 

\subsection{Fermi-LAT}
We extracted the GeV spectrum of \msp\ by analysing $\sim$89 months of \fermi-LAT data (latest version P8R2\footnote{http://fermi.gsfc.nasa.gov/ssc/}) from 2008 August 09
to 2016 December 31. 
All \fermi-LAT data reduction in this study was performed using the \fermi\ science tools v10r0p5 package.
We selected the data in a $10^{\circ}$ radius region of interest (ROI) centered at the 3FGL J2129.6--0427 position (R.A.,decl.) =$(322^{\circ}.439, -4^{\circ}.48664)$ with energies between 100 MeV and 300 GeV. We
collected all the events coverting both in the front and the back sections of the tracker (i.e., evtype = 3), and filtered the data with an event class (evtclass = 128) suitable for analysis of a point source. In addition, we selected photons with zenith angles smaller than $90^{\circ}$ to reduce the gamma-ray contamination caused by Earth albedo, and restricted data within time intervals determined as high quality (i.e., DATA\_QUAL $>$0). 

We performed a binned likelihood analysis using the \fermi\ science tool {\tt gtlike} in the region surrounding \msp.
For the background, we constructed a background emission model, including the Galactic diffuse emission (gll\_iem\_v06)
and the isotropic diffuse emission (iso\_P8R2\_SOURCE\_V6\_v06) distributed by the \emph{Fermi} Science Support Center, as well as all 3FGL catalogue sources (Acero et al. 2015) within $10^{\circ}$ from the centre of ROI to count for the spectral contribution. We note that the best-fit spectra for all of the background sources were obtained by eliminating all sources with $<3\sigma$.

\section{Data Analysis}

\subsection{Timing Analysis}

\msp\ has an orbital period of 15.2 hr that is seen in both X-ray (Hui et al. 2015b) and UV/optical (Hui et al. 2015b; Bellm et al. 2016). Previous \xmm\ observation showed a clear orbital modulation in the 0.3 to 10 keV band and a hint of hardness change over the orbital period (Hui et al. 2015b). In Figure 1, we show the \nustar\ ligt curves in difference energy bands as well as the hardness change using the radio timing ephemeris ($P_{orb}=0.63522741310$ days; $T_0= {\rm MJD}~55702.111161463$ where the companion inferior conjunction is at phase $\phi=0.25$). For comparison, we also plot the \xmm\ light curve. It is worth noting that the phase offset comparing to Hui et al. (2015b) is due to a different phase zero. In Hui et al. (2015b), the phase zero was arbitrarily chosen as the start of the \xmm\ observation.

The light curve profiles across the 0.3--40 keV band are very similar with minima and maxima at the orbital phase $\phi\sim0.1-0.3$ (companion inferior conjunction) and $\phi\sim0.6-0.9$ (companion superior conjunction), respectively. 
Hui et al. (2015b) found that in the \xmm\ data, the source has an indication of softening during the X-ray minimum. Owing to the limited photon statistic of \nustar, we do not see any significant hardness variation between 3--10 keV and 10--40 keV (Fig. 1). It is also likely due to the fact that the softening is due to a variation of the thermal emission ($kT\sim0.1-0.2$ keV) below 3 keV as indicated in spectral fits (Hui et al. 2015b; see also Section 3.2). Therefore, \nustar\ is not sensitive for the spectral change if the softening is due to a change in soft ($< 3$ keV) X-ray emission.
In Figure 1, a double-peaked feature at the X-ray maxima is seen in both \xmm\ and \nustar\ data and this feature was already reported for the \xmm\ data by Hui et al. (2015b). 

We also examined the orbital modulation and pulsation in the gamma-ray band. Using 8 years of \fermi-LAT data and a $1^{\circ}$ radius aperture, we did not find any significant orbital modulation and pulsation with an $H$ statistics of $\sim5$ and $\sim1$, respectively. This corresponds to a random probability of $\sim0.1$ and $\sim 0.7$ (de Jager et al. 2010). 

\begin{figure}
  \centering
  \epsfig{file=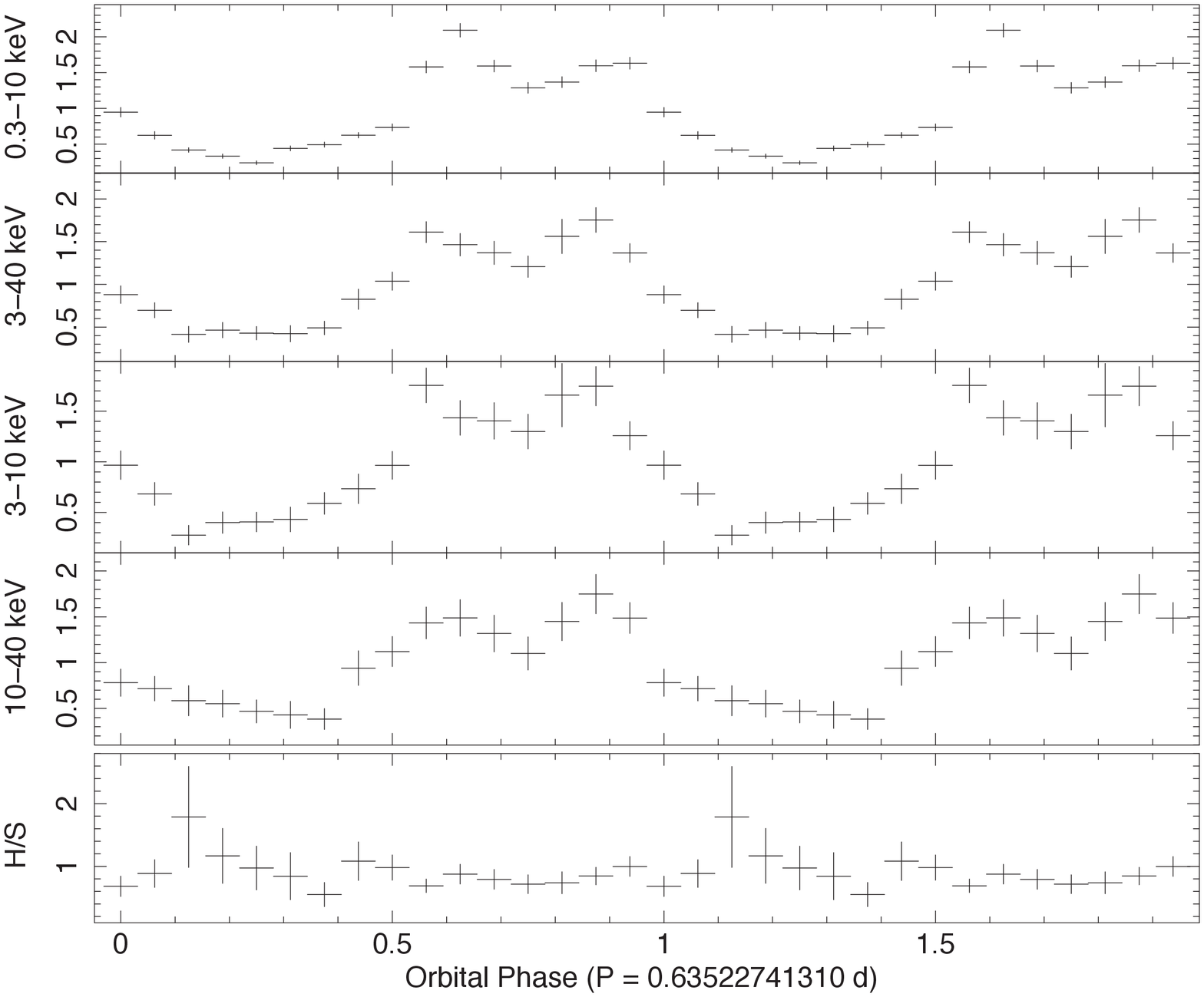,width=3.8in}
  \caption{Folded \xmm\ (0.3-10 keV) and \nustar\ light curves at different energy bands and hardness ratio (10-40 keV / 3-10 keV) variation of \msp\ with an orbital period of 0.63522741310 days. All light curves are in units of normalised count rates (background subtracted) and show the orbital modulation. On the other hand, the hardness ratio shows no sign of colour variation throughout the orbital period. The light curves are repeated twice for clarity.}
\end{figure}

\subsection{Spectral Analysis}

X-ray spectral fitting was performed with XSPEC version 12.9.1. All the quoted uncertainties are at the 90\% confidence level for one interesting parameter. Because \nustar\ data do not have sufficient photons to perform a meaningful phase-resolved spectroscopy, we will only focus on phase-averaged spectra in this paper. In order to increase the signal-to-noise of the dataset, we fit the two \nustar\ observations, each with spectra from FPMA and FPMB simultaneously. To constrain the soft X-ray spectral shape, we also fit the \nustar\ dataset simultaneously with the \xmm\ data. Although the \xmm\ observation and the two \nustar\ observations were all taken at different epoch, a quick-look of individual \nustar\ dataset revealed that there is no significant change in flux and spectral shape below 10 keV comparing to the \xmm\ observation.

We first started with a simple absorbed power-law model that was shown to be a reasonably good model for \xmm\ data (Hui et al. 2015b). To take cross-calibration between \nustar\ FPMA and FPMB and \xmm\ pn, MOS1/2 detectors, we also included a constant in all the spectral models discussed here. The X-ray emission is clearly seen up to about 40 keV without obvious emission and absorption features (see Fig. 2). Because the absorption is not well constrained (see Hui et al. 2015b) and it always converged to zero, we fixed it at a Galactic value of $4\times10^{20}$ cm$^{-2}$. This value is also consistent with the extinction derived from optical light curve modelling (Bellm et al. 2016). The best-fit photon index is $1.32\pm0.05$ with a $\chi^2 =155.78$ for 128 degrees of freedom (dof). 
The unabsorbed 0.3-79 keV flux is $1.22\times10^{-12}$ erg cm$^{-2}$ s$^{-1}$, corresponding to $2.86\times10^{32}$ erg s$^{-1}$. 
The spectral parameters are roughly consistent with previous \xmm\ observation (Hui et al. 2015b) but with additional data from \nustar, the fit quality of a simple power-law model is marginal. 

Since a neutron star in a binary system can emit detectable thermal emission, we next included a thermal component to build a composite model. We started by using a simple blackbody plus power-law model. A simple blackbody is a simple parametrization for  thermal emission and allows us to evaluate its possible contribution. While a simple power-law model yields an acceptable fit, a two-component model improves the fit and the additional blackbody component is statistically significant at 5$\sigma$ level by using an F-test. The best-fit blackbody temperature and photon index is $0.15\pm0.04$ keV and $1.16\pm0.08$, respectively. Assuming a source distance of 1.4 kpc, the blackbody emission radius is $0.27^{+0.24}_{-0.10}$ km and the blackbody emission contributes about 4\% of the total radiation in the 0.3-10 keV band. The thermal emission region is quite small comparing to the size of a typical neutron star. For such a small thermal emission region, it is likely from the hot polar cap of the MSP. To establish a more physical model, we replaced the blackbody component with a non-magnetic neutron star hydrogen atmosphere model ({\tt nsatmos} model in XSPEC; Heinke et al. 2006). We fixed the mass and size of the neutron star at $1.4M_\odot$ and 10 km, respectively. The derived effective temperature is $(9\pm3)\times10^5$ K and the emission radius is about $1.5^{+2.5}_{-0.7}$ km. The thermal contribution is about 7\% of the total emission in the 0.3-10 keV band. The unabsorbed 0.3-79 keV flux is $1.52\times10^{-12}$ erg cm$^{-2}$ s$^{-1}$, corresponding to $3.56\times10^{32}$ erg s$^{-1}$. Using a more massive neutron star ($1.74M_\odot$) as derived from optical light curve modelling (Bellm et al. 2016) does not change the spectral parameters significantly.
We summarize the best-fit parameters from the above spectral fits in Table 1.

\begin{figure}
  \centering
  \epsfig{file=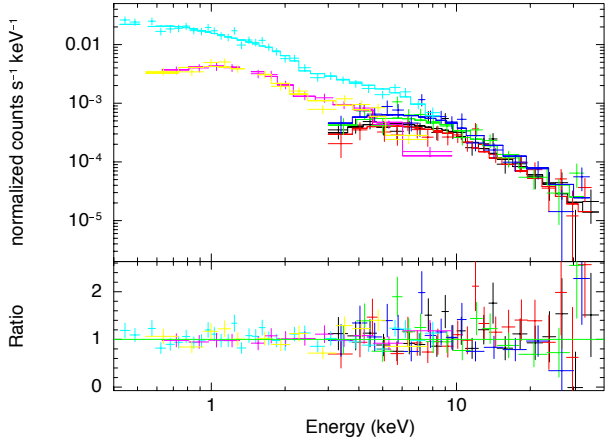,width=3.8in}
  \caption{Phase-averaged spectrum of \msp\ taken with \xmm\ (0.3-10 keV) and \nustar\ (3-40 keV). The best-fit spectrum (solid lines) is an absorbed power-law ($\Gamma=1.13$) plus a neutron star hydrogen atmosphere model ($T=9\times10^5$ K). }
\end{figure}

\begin{table}
\centering
\caption{Summary of Spectral Fits for \msp}
\begin{tabular}{cc}
\hline
\hline
Model & \\
\hline
Power-law & \\
\hline
$N_H$ ($10^{20}$ cm$^{-2}$) & 4.0 (fixed)\\
$\Gamma$ & $1.32\pm0.05$\\
$F_{0.3-10}$ ($10^{-13}$ erg cm$^{-2}$ s$^{-1}$) & $2.13^{+0.40}_{-0.07}$\\
$F_{3-79}$ ($10^{-13}$ erg cm$^{-2}$ s$^{-1}$) & $11.66^{+2.79}_{-1.34}$\\
$\chi^2_{\nu}/dof$ & 1.22/128\\
\hline
Power-law + Blackbody &\\
\hline
$N_H$ ($10^{20}$ cm$^{-2}$) & 4.0 (fixed)\\
$\Gamma$ & $1.16\pm0.08$\\
$kT$ (keV) & $0.15\pm0.04$\\
$R$ (km) & $0.27^{+0.24}_{-0.10}$\\
$F_{0.3-10}$ ($10^{-13}$ erg cm$^{-2}$ s$^{-1}$) & $2.24^{+0.12}_{-0.35}$\\
$F_{3-79}$ ($10^{-13}$ erg cm$^{-2}$ s$^{-1}$) & $14.37^{+2.12}_{-1.91}$\\
$\chi^2_{\nu}/dof$ & 1.0/126\\
\hline
Power-law + H atmosphere~$^a$ &\\
\hline
$N_H$ ($10^{20}$ cm$^{-2}$) & 4.0 (fixed)\\
$\Gamma$ & $1.13\pm0.08$\\
$T$ ($10^5$ K) & $9.0\pm3.0$\\
$R$ (km) & $1.5^{+2.5}_{-0.7}$\\
$F_{0.3-10}$ ($10^{-13}$ erg cm$^{-2}$ s$^{-1}$) & $2.10^{+0.23}_{-0.23}$\\
$F_{3-79}$ ($10^{-13}$ erg cm$^{-2}$ s$^{-1}$) & $14.86^{+2.25}_{-2.05}$\\
$\chi^2_{\nu}/dof$ & 0.98/126\\
\hline
\end{tabular}
\par
\smallskip
\begin{flushleft}
NOTE. - All fluxes are unabsorbed flux. \\
A distance of 1.4 kpc is assumed in all calculation.\\
$^a$~The neutron star's mass and size are fixed at $1.4M_\odot$ and 10 km, respectively.
\end{flushleft}
\end{table}

To obtain the GeV gamma-ray spectrum, we divided the \fermi-LAT energy range ($E>100$ MeV) into 8 segments.
The gamma-ray spectra of \msp\ are modeled using a power law with an exponential cutoff of the form
\begin{equation}
  \frac{dN}{dE} =N_{0} \left(\frac{E}{E_{0}} \right)
  ^{\Gamma}{\rm exp}\left[-\left(\frac{E}{E_{C}} \right)^b\right],
  \label{fit}
\end{equation}
where $N$ is the number of photons per unit time,
unit area and unit energy, $E$ is the energy of photon, $N_{0}$ is the normalization constant, $E_{0}$ is the scale factor of energy, $\Gamma$ is the spectral power-law index, and $E_{C}$ is the cut-off energy.
By fixing at $b=1$, we obtain the best fitting parameters as  $E_c=4810\pm1371$MeV and $\Gamma=1.7\pm0.12$, and its flux $F=(1.0\pm 0.1) \times 10^{-8}
\,{\rm photons~cm^{-2}~s^{-1}}$.

\subsection{Spatial Analysis}
\begin{figure}
  \centering
  \epsfig{file=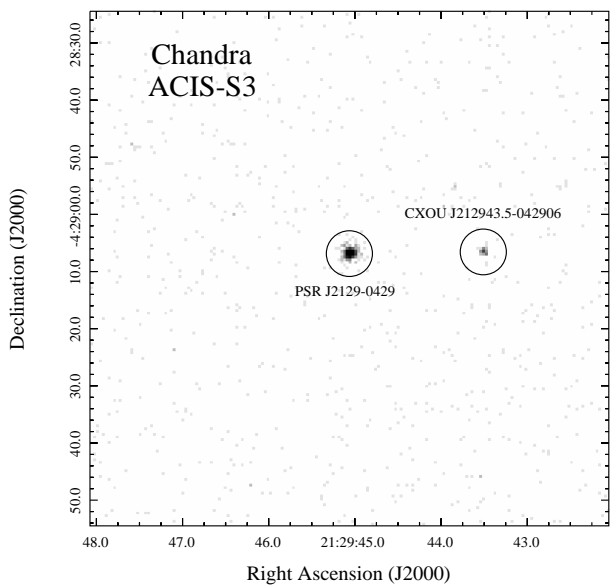,width=3.7in}
  \caption{\chandra\ 0.3-7 keV image of the region ($1.5^{'}\times1.5^{'}$) surrounding \msp. Apart from the MSP \msp\ seen as the brightest source in the image, a faint source toward west of the MSP is also detected.}
  \label{cxc}
\end{figure}

To investigate whether the apparently extended feature associated with \msp\ found by \emph{XMM-Newton} 
is originated from a PWN (Fig.~1 in Hui et al. 2015b), we analysed the image obtained by \emph{Chandra}. In Figure~\ref{cxc}, 
we show the ACIS-S3 image of the $1.5^{'}\times1.5^{'}$ field around \msp\ with a pixel size of $0.5^{"}\times0.5^{"}$.
Using a wavelet source detection algorithm (\emph{CIAO} tool {\tt wavdetect}), two sources are detected in this 
field. The brightest source (S/N=164$\sigma$) is detected at R.A. (J2000)=21:29:45.064, Dec (J2000)=-04:29:06.863 
with statistical uncertainties of $\delta$RA=$\delta$Dec=0.02 arcsec. This has an offset of 1.3 arcsec from 
the radio timing position of \msp\ (cf. ATNF pulsar catalogue, Manchester et al. 2005). But it is found to be 
consistent with the position of the optical counterpart within the tolerance of astrometric errors of both X-ray 
and optical observations (Bellm et al. 2016). 828 net counts are collected from an apeture of radius of 
$2.5^{"}$. The light curve of this source exhibits 
the same orbital modulation seen by \emph{XMM-Newton} and \emph{NuSTAR} (Figure~1).
For the spectrum, we found the best-fit spectral parameters and the flux are consistent with those we reported in Sec. 3.2 (see Table 1). Therefore, this bright source is undoubtably the X-ray counterpart of 
\msp. 

Another source is detected with a S/N ratio of $\sim19\sigma$ at 
R.A. (J2000)=21:29:43.510, Dec (J2000)=-04:29:06.567 ($\delta$RA=0.06, $\delta$Dec=0.07). 
It is designated as CXOU~J212943.5-042906 and locates $\sim23^{"}$ due west from \msp. Therefore, the putative feature identified by Hui et al. (2015b) 
is result of this previously unresolved source. 
At the position of CXOU~J212943.5-042906, we found one $r\sim21$ star in the Pan-STARRS catalogue.
Within a $2.5^{"}$ radius, there are 40 net counts collected from this source. Its spectrum 
can be well characterized by an absorbed power-law model with $N_{H}=1.1^{+6.0}_{-1.1}\times10^{21}$~cm$^{-2}$ and 
$\Gamma=1.9^{+0.9}_{-0.4}$ ($C$-stat.=2.56 for 5 d.o.f.). The unabsorbed flux in the $0.3-10$~keV band is found to be 
$F^{\rm unabs}_{0.3-10}=1.2\times10^{-14}$~erg~cm$^{-2}$~s$^{-1}$. It can also be modeled by a thermal plasma 
model with solar abundance (i.e XSPEC model {\tt mekal}), $N_{H}<4.7\times10^{21}$~cm$^{-2}$, 
$kT=6.7^{+21.5}_{-3.8}$~keV, $F^{\rm unabs}_{0.3-10}=1.1\times10^{-14}$~erg~cm$^{-2}$~s$^{-1}$ with 
similar goodness-of-fit ($C$-stat.=2.97 for 5 d.o.f.). Based on \chandra\ observations of serendipitous sources (e.g., Green et al. 2004), the X-ray flux and optical magnitude suggest that CXOU~J212943.5--042906 is very likely an AGN.

Since the X-ray photon statistic of CXOU~J212943.5-042906 is $<5\%$ of that of \msp, the problem of source 
confusion in the \emph{XMM-Newton} and \emph{NuSTAR} data will not alter the conclusions of the timing and spectral analysis 
presented in Sec. 3.1 and 3.2. Furthermore, this source is point like and isolated and we can now confirm that it is not associated with a PWN from \msp.

\section{Discussion}
By using \xmm\ , \nustar\ and \emph{Chandra} data, we have presented a broadband (0.3--40 keV) X-ray analysis of the redback MSP \msp. The source can be seen up to about 40 keV and the 0.64-day orbital period is clearly detected across the wavebands. The X-ray minimum and maximum correspond to the companion inferior conjunction and superior conjunction, respectively. These features are roughly associated with the two optical maxima seen in the UV/optical light curves (Hui et al. 2015b; Bellm et al. 2016). Moreover, a double-peaked structure is seen during the X-ray peak in all wavebands. All these are consistent with the results based on the 0.3--10 keV \xmm\ data (Hui et al. 2015b). With a high-resolution image taken with \chandra, \msp\ is now an isolated point source and we found no signature of a PWN.
The broadband X-ray spectrum can be described by a simple power-law model, and an additional thermal (simple blackbody or neutron star atmosphere) component can improve the fit significantly. We prefer the neutron star hydrogen atmosphere model over a blackbody because the emission region is more realistic ($\sim 1.5$ km in radius) and an atmosphere is expected above the surface of a MSP. The polar cap radius is given by the last closed magnetic field lines of a dipole, $r_{PC}=\sqrt{2\pi R_{NS}^3/P_s c}=1.6$ km, where the radius of the neutron star is assumed to be 10 km and the spin period $P_s$ is 7.62 ms. Our measurement of the polar cap size based on a neutron star atmosphere model is consistent with the theoretical value. In general, the broadband spectral fit is consistent with previous \xmm\ result (Hui et al. 2015b). With a better constraint of the non-thermal component provided by the \nustar\ data, we have shown that the thermal component is statistically required. The X-ray timing and spectral properties of \msp\ are very similar to those of the transitional MSP, PSR J1023+0038 (Bogdanov et al. 2011; Takata et al. 2014; Li et al. 2014). During the rotation-powered state, PSR J1023+0038 shows a similar two-component model and light curve profile.

Thanks to \nustar, hard X-rays from redback MSPs are seen in a couple of systems (Li et al. 2014; Kong et al. 2017) and they share some common properties. For example, they have a relatively hard ($\Gamma\sim1.2$) non-thermal power-law emission. In the case of PSR J1023+0038 and PSR J1723--2837, the X-ray emission extends all the way up to 79 keV. The X-ray emission shows orbital modulation with the X-ray minimum corresponds to the companion inferior conjunction. While the X-ray luminosity of \msp\ is about a factor of 2 fainter than that of PSR J1023+0038 and PSR J1723--2837, it shows a statistically significant thermal component presumably coming from the hot polar cap of the MSP. It is worth noting that the thermal component of PSR J1023+0038 is not formally required for the phase-averaged spectrum (Bogdanov et al. 2011).
The X-ray emission that modulates over the binary orbit can be explained by the synchrotron radiation of the pulsar wind particles
accelerated at the intra-binary shock as a result of interaction between pulsar wind and outflow from the companion. 

\begin{figure}
    \centering
\epsfig{file=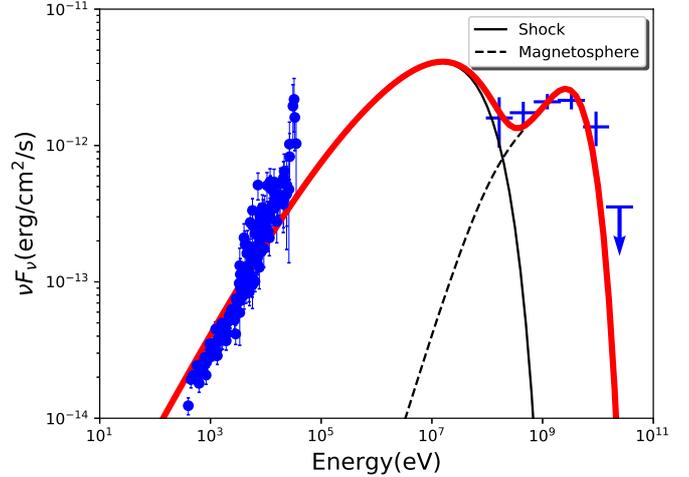,width=3.7in}
\caption{X-ray/gamma-ray spectrum of PSR~J2129--4029 (see Section 3.2 for data analysis). The dashed and solid black lines show the model spectra of the synchrotron
  emissions of the shocked pulsar wind particles and of the curvature radiation in the outer gap accelerator, respectively. We
  assume shock due to the interaction of the stellar wind and pulsar wind, for which the momentum ratio is $\eta_b=7$. In addition
  , the power-law index  of the shocked particles is assumed to be $p=1.5$. }
  \label{models}
\end{figure}

Owing to the similarity of PSR~J1023+0038 and \msp, we here use a similar intra-binary model to explain the behaviours of \msp.
In Li et al. (2014), we assumed
that the shock of PSR~J1023+0038 binary wraps the pulsar (Archibald et al. 2013), and we fit the X-ray data
with $\eta_b\sim 7$, where $\eta_b$ is the
momentum ratio of the stellar magnetic pressure and ram pressure of the pulsar wind, $\eta_b=B_*^2R_*^2c/L_{sd}\sim
5(B_*/10^2{\rm G})^2(R_*/2\times 10^{10}{\rm cm})^2(L_{sd}/5\times 10^{34}{\rm erg~s^{-1}})^{-1}$,
where $B_*$ and $R_*$ are  the stellar magnetic field and radius of the companion star, respectively. Because
of the modulation of the optical emission with the orbital phase, it is expected that
the spin of the companion star synchronizes with the orbital period ($P_{orb}\sim 0.64$d). With such
a rapidly spinning main-sequence star, the stellar dynamo process could  generate a stellar
magnetic field of several kG (Reiners et al. 2009), and the magnetic pressure could overcome the
pulsar wind pressure. To model the X-ray light curve of PSR~J2129--0429,
we also apply the momentum ratio $\eta_b=7$, and calculate the shock geometry with the method
discussed by Canto et al. (1996).  We assume that 
an isotropic pulsar wind that carries the spin down power and that wind energy at the shock
is carried by the electrons and positrons. In our calculation, we apply
the magnetization parameter $\sigma=0.1$ for
the ratio of the magnetic energy and kinetic energy of the pulsar wind at the shock.
The pulsar wind is compressed by the shock, and the wind particles are accelerated to higher energies. We assume that the distribution of the
accelerated particles is described by a power law function, and we determine
the maximum Lorentz factor  by  balancing between
the synchrotron loss time scale and the acceleration time scale, yielding the maximum Lorentz factor $\gamma_{max}=[9m_e^2c^4/(4e^3B_s)]^{1/2}$, where $B_s$ represent the magnetic field strength just after the shock. 
We assume the minimum Lorentz factor is comparable to the Lorentz factor of the bulk motion in the up-stream
region, which corresponds to $\gamma_{pw}\sim 10^{3-6}$ depending on the spin down parameters (Takata et al. 2017).
In this paper, we apply $\gamma_{min}=10^4$. We note that the results of the our calculation is insensitive
to the minimum Lorentz factor.  We obtain the magnetic field in the shocked pulsar wind at the shock  from $B_s=3\sqrt{L_{sd}\sigma/[r_s^2c(1+\sigma)]}$, and in the down stream region, we
estimate the magnetic field strength  at the distance $r$  from the shock using
the magnetic flux conservation, that is, $B(r)r=$constant along the flow.
In the downstream region, the particles lose their energy via the adiabatic loss and synchrotron loss. For GeV band, we expect that the emission is originated from the pulsar magnetosphere, and fit the observation with a
outer gap model developed by Wang et al. (2010).
Figure~\ref{models} shows the results of our model fitting for X-ray and GeV gamma-ray bands. Here, we assume the accelerated particles at the shock follow a power-law distribution with $p=1.5$.

\begin{figure}
  \centering
  \epsfig{file=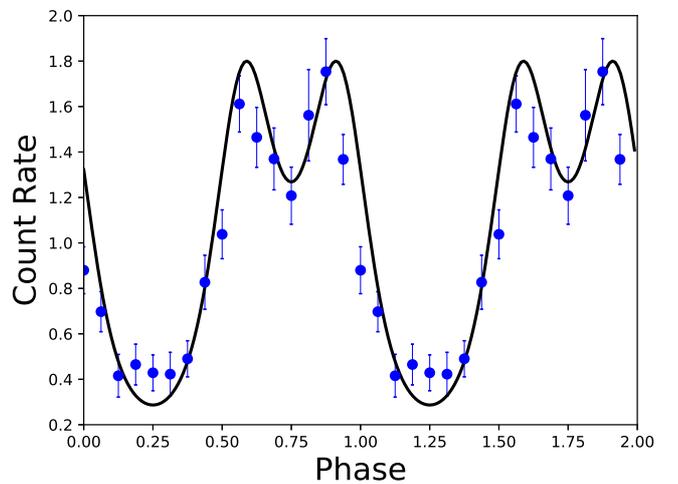,width=3.7in}
  \caption{Orbital modulation (in units of normalised count rate) of PSR~J2129--4029 in the 3--40 keV band. Phase 0.25 corresponds to the companion inferior conjunction,
    where the companion is located between the pulsar and Earth. In the present model (solid line),
    the Doppler boosting causes the orbital modulation. We assume  $\eta_b=7$ as the momentum ratio, $v\sim 0.6c$ as the velocity of the shocked pulsar wind, and an inclination of $\sim70^{\circ}$.}
  \label{modell}
\end{figure}

In our model, the shock wraps the pulsar and because the shocked pulsar wind has a finite velocity, the X-ray orbital modulation can be explained by Doppler boosting. To calculate the model light curve, we assume that the shocked pulsar wind flows along the shock cone with a constant speed. We further assume the velocity of the shocked pulsar wind as $v=0.6 c$ and a system's inclination of $\theta_s\sim 70^{\circ}$.
In Figure~\ref{modell}, we compare
the model light curve and the \nustar\ 3--40 keV observations, and the model can reproduce all the major features of the observations. The peaks in the model light curve correspond
to the positions when the line of sight cuts through the
surface of the shock cone. It is worth noting that from optical light curve modelling, the derived inclination is $80.5\pm7^{\circ}$ (Bellm et al. 2016), implying a nearly edge-on system.

To compare with other two redback MSPs, PSR\,J1023+0038 and PSR\,J1723--2837, that have \nustar\ detections up to 79 keV (Li et al. 2014; Kong et al. 2017), \msp\ shares several common behaviors. All three redbacks have a relatively hard non-thermal power-law spectrum with a photon index of $\sim 1.2$ and they all show X-ray orbital modulation. This non-thermal emission as well as the X-ray modulation can be modelled with an intrabinary shock model. Perhaps the unique features of \msp\ is the thermal emission from the neutron star on top of the dominant power-law emission and the double-peaked X-ray light curve. The thermal emission of \msp\ could be the result of a slightly higher temperature of the polar cap and/or geometrical effect. The double-peaked structure of the X-ray light curve can be explained naturally by a shock warping pulsar with Doppler boosting effect in a nearly edge-on ($\sim70^{\circ}$) orbit. For PSR\,J1023+0038 and PSR\,J1723--2837, their inclination is about $40^{\circ}$ (Bogdanov et al. 2011; Crawford et al. 2013), making the double-peaked feature less apparent (see Figure 6 in Li et al. 2014). Lastly, \msp\ has the longest binary orbital period comparing to the other two and that may also account the differences.

\section*{Acknowledgements}
A.K.H.K. is supported by the Ministry of Science and Technology of the
Republic of China (Taiwan) through grants 105-2119-M-007-028-MY3 and
105-2112-M-007-033-MY2. J.T. is supported by the National Science Foundation of China (NSFC) grants of Chinese Government under 11573010, U1631103, and 11661161010. C.Y.H. is supported by the National Research Foundation of Korea through grant 2016R1A5A1013277. P.H.T. is supported by the NSFC grants 11633007 and 11661161010.
Support for this work was partially provided by the National Aeronautics and Space Administration through Chandra Award Number GO7-18036X issued by the Chandra X-ray Observatory Center, which is operated by the Smithsonian Astrophysical Observatory for and on behalf of the National Aeronautics Space Administration under contract NAS8-03060.

\label{lastpage}
\end{document}